\documentclass[conference]{IEEEtran}
\IEEEoverridecommandlockouts

\usepackage{cite}
\usepackage{amsmath,amssymb,amsfonts}
\usepackage{algorithmic}
\usepackage{graphicx}
\usepackage{times}
\usepackage{courier}
\usepackage{bm}
\usepackage{times}
\usepackage{url}
\usepackage[small]{caption}
\usepackage{graphicx}
\usepackage{booktabs}
\usepackage{amsmath}
\usepackage{amsthm}
\usepackage{algorithm}
\usepackage{algorithmic}
\usepackage[switch]{lineno}
\def\BibTeX{{\rm B\kern-.05em{\sc i\kern-.025em b}\kern-.08em
    T\kern-.1667em\lower.7ex\hbox{E}\kern-.125emX}}
\begin{document}

\title{Challenges and Opportunities in 3D Content Generation\\

}

\author{\IEEEauthorblockN{Ke Zhao, Andreas Larsen}
kezhao421@gmail.com \\

}

\maketitle

\begin{abstract}
This paper explores the burgeoning field of 3D content generation within the landscape of Artificial Intelligence Generated Content (AIGC) and large-scale models. It investigates innovative methods like Text-to-3D and Image-to-3D, which translate text or images into 3D objects, reshaping our understanding of virtual and real-world simulations. Despite significant advancements in text and image generation, automatic 3D content generation remains nascent. This paper emphasizes the urgency for further research in this area. By leveraging pre-trained diffusion models, which have demonstrated prowess in high-fidelity image generation, this paper aims to summary 3D content creation, addressing challenges such as data scarcity and computational resource limitations. Additionally, this paper discusses the challenges and proposes solutions for using pre-trained diffusion models in 3D content generation. By synthesizing relevant research and outlining future directions, this study contributes to advancing the field of 3D content generation amidst the proliferation of large-scale AIGC models.
\end{abstract}

\begin{IEEEkeywords}
3D generation, Diffusion models
\end{IEEEkeywords}

\section{Introduction}
In the wave of Artificial Intelligence Generated Content (AIGC) and large-scale models~\cite{rombach2022high,saharia2022photorealistic,sanghi2022clip}, creating three-dimensional (3D) content has become an exciting research area, involving various innovative methods like Text-to-3D~\cite{nichol2022point,jun2023shap,lin2023magic3d,qian2023magic123,wang2023prolificdreamer} and Image-to-3D~\cite{shi2023zero123++,liu2023zero,jain2022zero,raj2023dreambooth3d}, translating text or images into 3D objects. These advancements are reshaping our understanding of building virtual worlds and simulating real ones.

In recent years, with the continuous improvement of deep learning technology, significant achievements have been made in text and image generation within the AIGC field. Text generation techniques have been widely applied in language translation, question answering systems, and other language-related tasks, while image generation techniques can create visual images based on textual descriptions. However, compared to text and image generation, automatic generation of 3D content is still in its early stages and has not achieved the same level of breakthrough.

Given this, further research and development in 3D content generation technology are urgently needed. 3D content finds broad applications in fields such as filmmaking, architectural design, virtual reality (VR), and augmented reality (AR). Currently, the mainstream production of 3D content still relies on professional 3D modelers, limiting production efficiency and raising industry entry barriers. Therefore, utilizing artificial intelligence technology to automate 3D content generation can significantly enhance production efficiency, reduce industry barriers, and further drive related industries' development.

In the field of 3D content generation, the application of zero-shot learning methods is relatively limited, mainly due to the smaller scale and diversity of 3D data compared to 2D image data. Traditional 3D content generation methods typically rely on specialized 3D datasets to train generative models. While these methods can generate 3D objects with consistent geometric structures, they are limited by the lack of large 3D datasets, the absence of efficient 3D content generation architectures, and the enormous computational resources required to train these models. These factors collectively make it challenging for traditional 3D content generation methods to make significant progress in the short term. Therefore, this study will focus on how to use pre-trained diffusion models to guide 3D content generation.

As a powerful generative model, diffusion models have demonstrated their capabilities in training on billions of image-text pairs, particularly showing the ability to high-fidelity images generation and editing~\cite{li2023layerdiffusion,hertz2022prompt,ruiz2023dreambooth,li2024tuning,brooks2023instructpix2pix,couairon2022diffedit,gafni2022make,ramesh2021zero}. Applying pre-trained diffusion models to 3D content generation can significantly reduce the demand for computational resources and the reliance on 3D datasets, thereby enhancing the feasibility and efficiency of 3D content generation. This paper will delve into and analyze methods for 3D content generation, with a special focus on the prior knowledge of diffusion models and the representation of 3D data. 3D content generation tasks are mainly divided into two categories: Text-to-3D and Image-to-3D. To comprehensively evaluate the advantages and limitations of different methods, this study will conduct a horizontal comparison of these models in terms of efficiency and quality. Additionally, this paper will discuss the challenges faced when using pre-trained diffusion models for 3D content generation and propose potential solutions.

To further expand the depth and breadth of this research, it is necessary to first review the technical background and theoretical foundation of 3D content generation, including traditional methods of 3D modeling, basic concepts of 3D graphics, and the latest developments in the AIGC field. We will then detail the working principles of diffusion models and how they are trained to handle image-text pairs, as well as discuss how these models can adapt to 3D content generation tasks. This paper will subsequently describe in detail the specific implementation methods of Text-to-3D and Image-to-3D, including data preprocessing, networks, loss functions, etc. Faced with challenges in 3D content generation, such as data scarcity, computational resource limitations, and the diversity of generated content, we will discuss their feasibility and potential impact.

The contributions of this study can be summarized as follows:
\begin{itemize}
 \item This paper summarizes relevant research on 3D content generation and analyzes the specific implementation of text-to-3D and image-to-3D, discussing the challenges faced when using diffusion models for 3D content generation.
 \item This paper outlines potential future research directions in the field of 3D content generation under the backdrop of large-scale AIGC models.
\end{itemize}

\section{Related work}

\subsection{3D Representation}
3D Representation is a crucial aspect of computer graphics, enabling the digital visualization and manipulation of three-dimensional objects and scenes. Different methods of 3D representation offer varying degrees of detail, efficiency, and application suitability. Here, we delve into several key types of 3D representations, including Explicit Representations, Point Clouds, and Voxels.

\paragraph{Explicit Representations}

Explicit representations define 3D objects with precise mathematical descriptions. This category includes mesh-based methods, which represent surfaces using vertices, edges, and faces. Meshes are widely used due to their flexibility and efficiency in rendering.

\paragraph{Meshes}
A mesh is composed of polygons, typically triangles, which are connected by their edges and vertices to form the surface of a 3D object. Meshes are favored in applications requiring detailed and smooth surfaces, such as video games, animations, and simulations. They allow for efficient rendering~\cite{botsch2010polygon,shirman1987local} and easy manipulation but can become complex when representing highly detailed surfaces.

\paragraph{Point Clouds}

Point clouds~\cite{aliev2020neural,wiles2020synsin,kopanas2021point,kopanas2021point} represent 3D shapes as a collection of discrete points in space. Each point has a specific position (x, y, z coordinates) and often includes additional attributes like color or normal vectors. Point clouds are particularly useful in scanning and reconstruction applications, as they can directly capture the surface geometry of objects from real-world scans. They are used in fields like autonomous driving, where LiDAR sensors generate point cloud data to represent the environment. However, point clouds can be sparse and may require significant processing to convert into other forms of representation, like meshes or voxel grids, for further use. The lack of explicit connectivity between points can also complicate certain operations like surface reconstruction and rendering.

\paragraph{Voxels}

Voxels~\cite{liu2020neural, fridovich2022plenoxels, schwarz2022voxgraf, sun2022direct,maturana2015voxnet}, or volumetric pixels, are the 3D equivalent of pixels in a 2D image. They divide the 3D space into a regular grid of small cubes, each storing information about the material present at that location. Voxel grids are highly suitable for applications requiring a uniform representation of space, such as medical imaging (CT and MRI scans), geological modeling, and some types of simulations (e.g., fluid dynamics). They are excellent for representing volumetric data and can easily handle complex topologies and internal structures of object. The primary challenge with voxels is their memory consumption, as high-resolution voxel grids can become extremely large. This makes them less efficient for detailed surface representations compared to meshes. However, advancements in sparse voxel representations and efficient storage techniques are mitigating these issues.

\subsection{Implicit Representations}
\paragraph{Neural Radiance Fields}
Neural Radiance Fields (NeRF)~\cite{mildenhall2020nerf, niemeyer2021giraffe,barron2021mip,barron2022mip, verbin2022ref,li2023dynibar,chen2021mvsnerf} is a groundbreaking technique in computer vision that synthesizes novel views of complex 3D scenes from a sparse set of 2D images. NeRF represents a scene using a continuous 5D function that maps spatial coordinates and viewing directions to color and density values. This enables the creation of photorealistic images from new viewpoints by interpolating between the input images. NeRF employs a fully-connected neural network to optimize a volumetric scene function, which, through volume rendering techniques, accumulates colors and densities to form a 2D image. The process involves marching rays through the scene, sampling 3D points, and predicting color and density values for each point using the neural network. This allows for high-fidelity reconstructions of scenes with intricate details and realistic lighting effects. Despite its impressive capabilities, the classic NeRF approach has some limitations, such as long training times and the requirement to train a new model for each scene. To address these challenges, several extensions and improvements have been proposed. For instance, InstantNeRF~\cite{muller2022instant} utilizes a multiresolution hash table to accelerate training, while PixelNeRF~\cite{yu2021pixelnerf} leverages convolutional neural networks to enable view synthesis from a single image without explicit 3D supervision.

\paragraph{3D Gaussian Splatting}
3D Gaussian Splatting~\cite{abdal2023gaussian,chen2023text,tang2023dreamgaussian,liu2023humangaussian,yi2023gaussiandreamer} is an advanced rasterization technique used for real-time rendering of photorealistic scenes. It begins by estimating a point cloud from a set of 2D images using a method called Structure from Motion (SfM). Each point in this cloud is then converted into a 3D Gaussian, characterized by its position, color, transparency, and shape. These Gaussians are then rasterized to create high-quality images that can be rendered in real-time. The main advantage of 3D Gaussian Splatting over traditional methods like Neural Radiance Fields (NeRF) and photogrammetry lies in its efficiency and real-time capabilities. Unlike NeRF, which can be computationally intensive and slow to render, Gaussian Splatting offers faster rendering speeds while maintaining high visual quality. This makes it particularly suitable for applications requiring real-time interactions, such as virtual reality (VR) and augmented reality (AR). The training process for 3D Gaussian Splatting involves rasterizing the Gaussians, calculating the loss based on the difference between the rasterized image and the ground truth, and adjusting the Gaussian parameters to minimize this loss. This process allows for detailed and accurate representation of complex scenes with high efficiency. Despite its advantages, 3D Gaussian Splatting does require significant VRAM and is not yet fully compatible with existing rendering pipelines. However, its potential applications are vast, ranging from enhancing virtual property tours and urban planning to creating photorealistic avatars for telepresence in VR environments.

\subsection{Diffusion models}

Diffusion models~\cite{ho2020denoising,rombach2022ldm,bai2023high,luo2021diffusion,cai2020learning} are based on a Markov chain that transitions data points from a clear state to a diffused state and vice versa. The core idea is to learn the reverse process of this Markov chain, which allows for sampling from complex distributions.

\paragraph{Forward Process (Diffusion)}

The forward process can be thought of as a Markov chain with transition probabilities \( p(x_{t-1} | x_t) \), where \( x_t \) is the data at step \( t \) and \( x_{t-1} \) is the data at step \( t-1 \). The transition is typically defined by adding Gaussian noise:
\begin{equation}
x_{t-1} = f(x_t) + \epsilon_t    
\end{equation}

where \( f(x_t) \) is the reverse process (denoising), and \( \epsilon_t \) is Gaussian noise with variance \( \sigma_t^2 \). The variance \( \sigma_t \) is typically chosen from a predefined schedule.

\paragraph{Reverse Process (Denoising)}

The reverse process involves learning a model \( f_{\theta}(x_t, t) \) that, given a noisy data point \( x_t \) and the step \( t \), predicts the previous data point \( x_{t-1} \). This is achieved by training the model to minimize the following loss function:

\begin{equation}
L(\theta) = \mathbb{E}_{x_0 \sim p(x_0)} \left[ \| x_0 - f_{\theta}(x_T, T) \|_2^2 \right]
\end{equation}

where \( x_0 \) is the original data point, \( x_T \) is the fully diffused data point after \( T \) steps, and \( p(x_0) \) is the data distribution.

\paragraph{Training Objective}

The training objective can also be formulated in terms of the log-likelihood of the data under the diffusion process. The joint distribution of the data and the noise is given by:

\begin{equation}
p(x_0, x_1, ..., x_T) = p(x_0) \prod_{t=1}^T p(x_{t-1} | x_t)
\end{equation}

The log-likelihood of \( x_0 \) can be written as:
\begin{equation}
\log p(x_0) = \log \left( \prod_{t=1}^T p(x_{t-1} | x_t) \right) + \log p(x_0)
\end{equation}

By introducing the reverse process, the log-likelihood can be maximized by minimizing the negative log-likelihood:
\begin{equation}
-\log p(x_0) = -\log \left( \prod_{t=1}^T p(x_{t-1} | x_t) \right) - \log p(x_0) + C
\end{equation}

where \( C \) is a constant term that does not depend on \( \theta \).

\paragraph{Sampling}

For sampling, the model is used in reverse, starting from a noise distribution:
\begin{equation}
x_{t-1} = f_{\theta}(x_t, t)
\end{equation}

By iteratively applying \( f_{\theta} \) and decreasing \( t \), one can generate new samples that approximate the original data distribution \( p(x_0) \).

In fact, the variance schedule \( \sigma_t \) can be chosen in various ways, such as a linear or exponential schedule. The denoising model \( f_{\theta} \) can be any neural network architecture, often a deep convolutional network. The training process involves optimizing \( \theta \) to minimize the reconstruction loss or maximize the log-likelihood.

In this section, we provides a more in-depth look at the diffusion model's principles and the mathematical framework behind it. The actual implementation may involve additional complexities and optimizations.

\section{3D Content generation}
Diffusion models have significant advantages in generating 3D content, especially when combined with NeRF (Neural Radiance Fields) and 3D Gaussian Splatting technologies. NeRF represents scenes as continuous volumetric fields, enabling the generation of realistic 3D scene renderings. Diffusion models, on the other hand, provide an effective method for generating high-quality 3D models from noise. By combining these two technologies, the details and realism of 3D assets can be further enhanced. The 3D Gaussian Splatting technology is a method of representing 3D data as a set of Gaussian distributions, which helps maintain the structural integrity of the 3D models during the diffusion process. By integrating diffusion models with these technologies, researchers can explore new methods of generating 3D assets that not only improve generation efficiency but also provide greater flexibility and control while preserving details.

\begin{figure}[t]
	\centering

	\includegraphics[width=\linewidth]{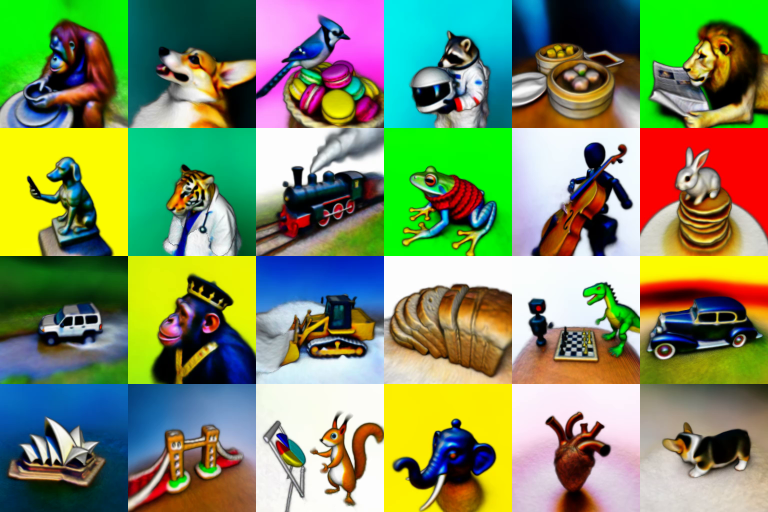}

	\caption{DreamFusion~\cite{poole2022dreamfusion} utilizes a pretrained text-to-image diffusion model to generate realistic 3D models from text prompts..}
	\label{fig:svd5}
\end{figure}

\begin{figure}[t]
	\centering

	\includegraphics[width=\linewidth]{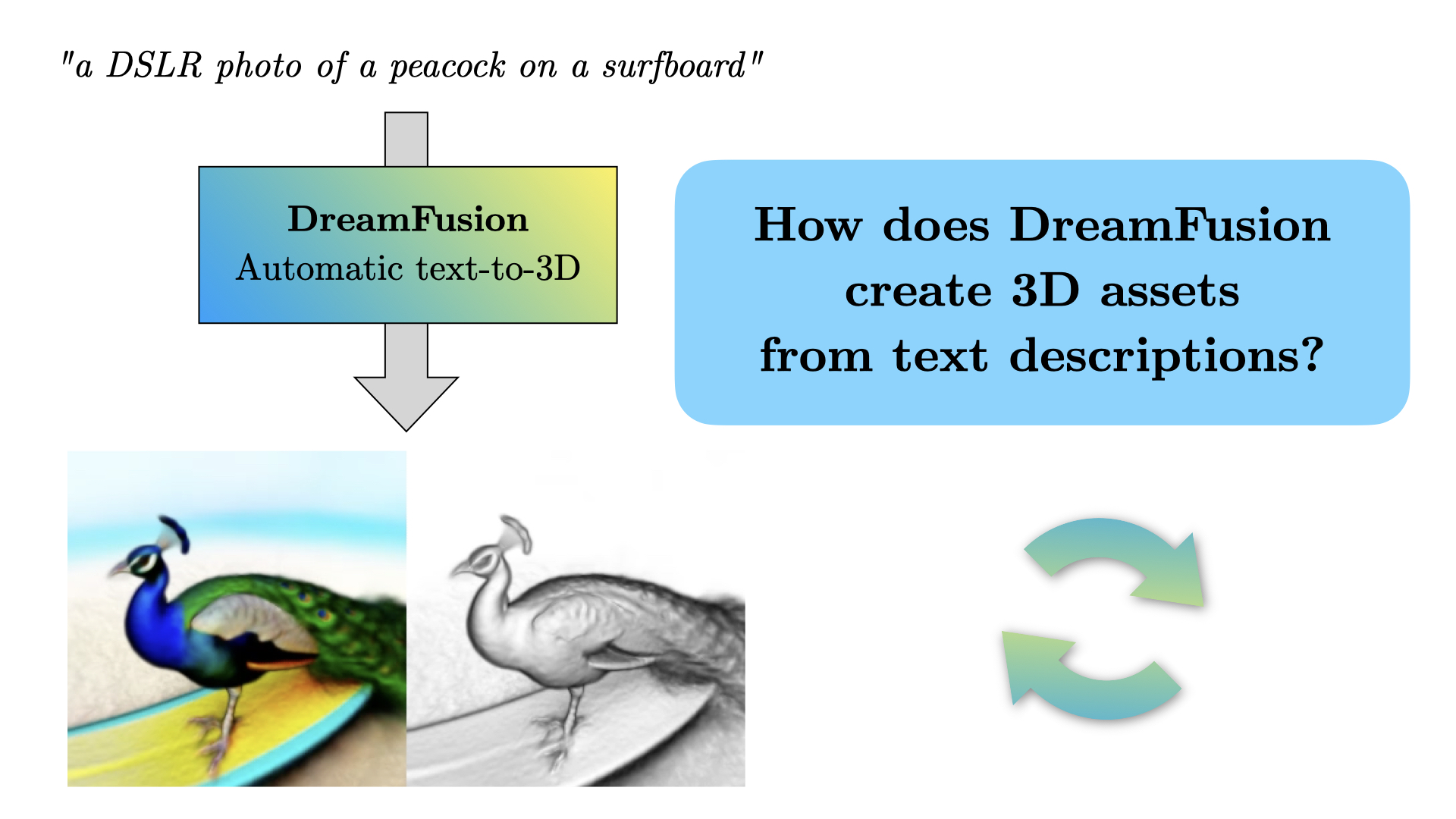}

	\caption{We demonstrate DreamFusion's capability to translate text into 3D assets.}
	\label{fig:03}
\end{figure}

\begin{figure}[t]
	\centering

	\includegraphics[width=\linewidth]{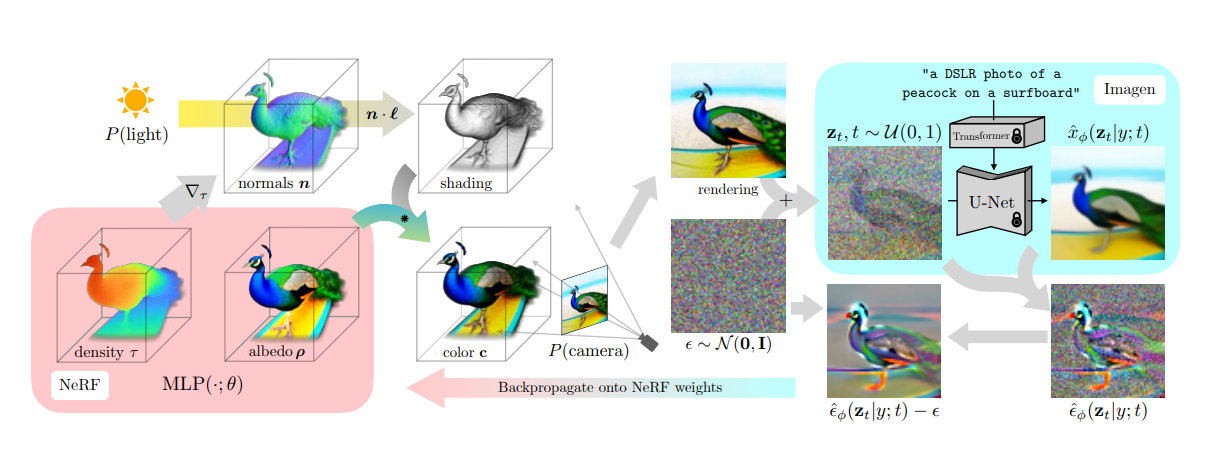}

	\caption{We show the pipeline of Dreamfusion.}
	\label{fig:03}
\end{figure}

\subsection{Text-to-3D}

Text-to-3D~\cite{li2024art3d,poole2022dreamfusion,lin2023magic3d,li2024generating,chan2023generative,tang2023make,shen2021deep,wang2023prolificdreamer,li2023sweetdreamer,metzer2023latent} is an emerging interdisciplinary field that focuses on converting textual descriptions into three-dimensional models, such as shown in Fig. 1, 2 and 3.. This technology has a wide range of applications, including virtual reality, gaming, and design. The goal is to create a system that can interpret natural language and generate a 3D representation that aligns with the textual description.

Data Representation:
The core of text-to-3D methods lies in how they represent and process data. There are primarily two approaches:

Volumetric Representation: Early methods used voxel grids to represent 3D shapes. This approach allows for direct 3D shape manipulation but can be computationally intensive and storage-heavy.
Surface Representation: More recent methods focus on mesh surfaces, which are more memory-efficient and align better with human perception of 3D shapes.
Neural Networks and Learning:
The evolution of neural networks has been pivotal in advancing text-to-3D conversion. Key developments include:

Convolutional Neural Networks (CNNs): Used for feature extraction from images, which can be used to guide the generation process.
Recurrent Neural Networks (RNNs): Effective for sequential data processing, such as sentences in natural language.
Transformers: State-of-the-art models that excel at handling sequential data and have been adapted for text-to-3D tasks.
Generative Models:
Generative models are at the heart of text-to-3D systems. They include:

Generative Adversarial Networks (GANs)~\cite{goodfellow2014generative,abdal2019image2stylegan}: Known for their ability to generate realistic images, some adaptations are used to generate 3D shapes.
Variational Autoencoders (VAEs)~\cite{vae,kusner2017grammar,pu2016variational}: Used to learn a low-dimensional representation of the data, which can then be decoded into 3D shapes.
Diffusion Models: A newer approach that has shown promise in generating high-quality 3D shapes from text.
Attention Mechanisms:
Attention mechanisms~\cite{vaswani2017attention} have proven to be highly effective in text-to-3D tasks by allowing the model to focus on different parts of the input text and generate corresponding 3D features.

Depth priors: Depth estimation~\cite{li2023efficient,ranftl2020towards,bhat2021adabins,guizilini20203d} provides a prior for depth information, supervising subsequent model generation. The entire process begins with initializing a 3D model, followed by rendering images with added Gaussian noise from random angles. Subsequently, the diffusion model~\cite{bai2023high} is utilized to optimize the 3D model through backpropagation using SDS loss and a series of reference image losses.

Loss Functions:
Custom loss functions are crucial for training models to generate accurate 3D shapes. Common loss functions include:

Perceptual Loss: Measures the difference between the generated shape and a target shape based on perceptual features.
Chamfer Distance: A metric that measures the similarity between two point clouds, often used to compare the generated mesh with a reference mesh.
Earth Mover's Distance (EMD): Used to measure the distribution difference between two sets of points, which can be applied to 3D shapes.
Evaluation Metrics:
Evaluating the quality of generated 3D models is critical. Metrics used include:

Fidelity to Description: How well the generated model matches the textual description.
Visual Quality: The realism and detail of the generated 3D shapes.
Diversity: The range of shapes that can be generated from different textual inputs.
Applications:
The applications of text-to-3D are vast and include:

Content Creation: Automatically generating 3D content for games, films, or virtual environments.
Design Assistance: Assisting designers by quickly converting textual ideas into visual 3D concepts.
Educational Tools: Teaching 3D modeling concepts through natural language interactions.
Challenges:
Despite the progress, text-to-3D faces several challenges:

Ambiguity in Language: Textual descriptions can be ambiguous, making it difficult to generate a single, accurate 3D model.
Complexity of 3D Shapes: Capturing the full complexity of 3D shapes from text is a complex task.
Computational Resources: High-quality 3D generation can be computationally expensive.

\subsection{Image-to-3D}

Recent research has been focusing on exploring image-to-3D techniques~\cite{jiang2022nerffaceediting,lin2023magic3d,liu2023meshdiffusion,li2023archi,liu2023zero,sun2023dreamcraft3d,melas2023realfusion} to describe scene details and appearance more intuitively. These methods aim to reconstruct outstanding and high-fidelity 3D models from given images, serving as the primary means to depict the visual effects of scenes, which are more detailed than language. Similar to text-to-3D methods, several image-to-3D methods utilize volume representations like NeRF to naturally introduce multi-view consistency. For example, NeuralLift-360\cite{xu2023neurallift} utilizes monocular depth and CLIP-guided diffusion priors to respectively regulate geometry and appearance optimization, thereby elevating a single image to a 3D scene represented by NeRF. RealFusion\cite{melas2023realfusion} and NeRDi\cite{deng2023nerdi} extract text embeddings via text inversion to fine-tune pre-trained image diffusion models and combine score distillation loss to optimize volume representations. Magic123\cite{qian2023magic123} introduces the 3D prior of the pre-trained viewpoint-conditioned diffusion model Zero-1-to-3\cite{liu2023zero}(as shown in Fig. 4 and 5.) in a coarse-to-fine framework through two optimization stages to generate texture meshes matching specified images. Make-it-3D\cite{tang2023make} enhances texture and geometric structure in the refinement stage of a two-stage method to produce high-quality textured point clouds. Subsequent works\cite{sun2023dreamcraft3d, yu2023hifi} have been continuously improving upon previous results. Recently, 3D Gaussian Splatter (3DGS)\cite{kerbl20233d} has emerged as a promising modeling and real-time rendering technique. Based on 3DGS, DreamGaussian\cite{tang2023dreamgaussian} proposes an efficient two-stage framework for text-driven and image-driven 3D generation. In the first stage, DreamGaussian utilizes SDS loss (i.e., 2D diffusion prior and CLIP-guided diffusion prior) to generate the target object represented by 3D Gaussians. Subsequently, DreamGaussian~\cite{tang2023dreamgaussian} extracts textured meshes from the optimized 3D Gaussians by querying local densities and refines textures in UV space. To evaluate viewpoint synthesis quality, CLIP similarity, PSNR, and LPIPS are commonly used as metrics.

PSNR: PSNR is a widely used standard for quantifying image reconstruction or image compression quality. It measures the pixel-level difference between the original image and the compressed or reconstructed image. PSNR is computed based on the Mean Squared Error (MSE) between two images. Generally, a higher PSNR value indicates a closer quality match between the reconstructed image and the original image. It primarily assesses pixel-level similarity between the reconstructed or compressed image and the original image, but may not always align with human perceptual differences.

LPIPS: LPIPS is a more modern deep learning-based metric used to evaluate the perceptual quality and similarity of images. LPIPS computes similarity by comparing activations of deep neural networks when processing two images. This approach aims to be closer to the human visual perception system. LPIPS is used to assess perceptual similarity of images, especially in cases where pixel-level metrics might not capture all aspects of human perception.

CLIP-Similarity: CLIP-Similarity is a metric used to evaluate semantic similarity of images based on features extracted by the CLIP model. Unlike traditional image similarity metrics that focus on pixel-level details, CLIP-Similarity measures the semantic or contextual similarity between two images. It is particularly useful when evaluating criteria beyond purely visual or pixel-level accuracy and involves contextual and conceptual consistency.

\begin{figure}[t]
	\centering

	\includegraphics[width=\linewidth]{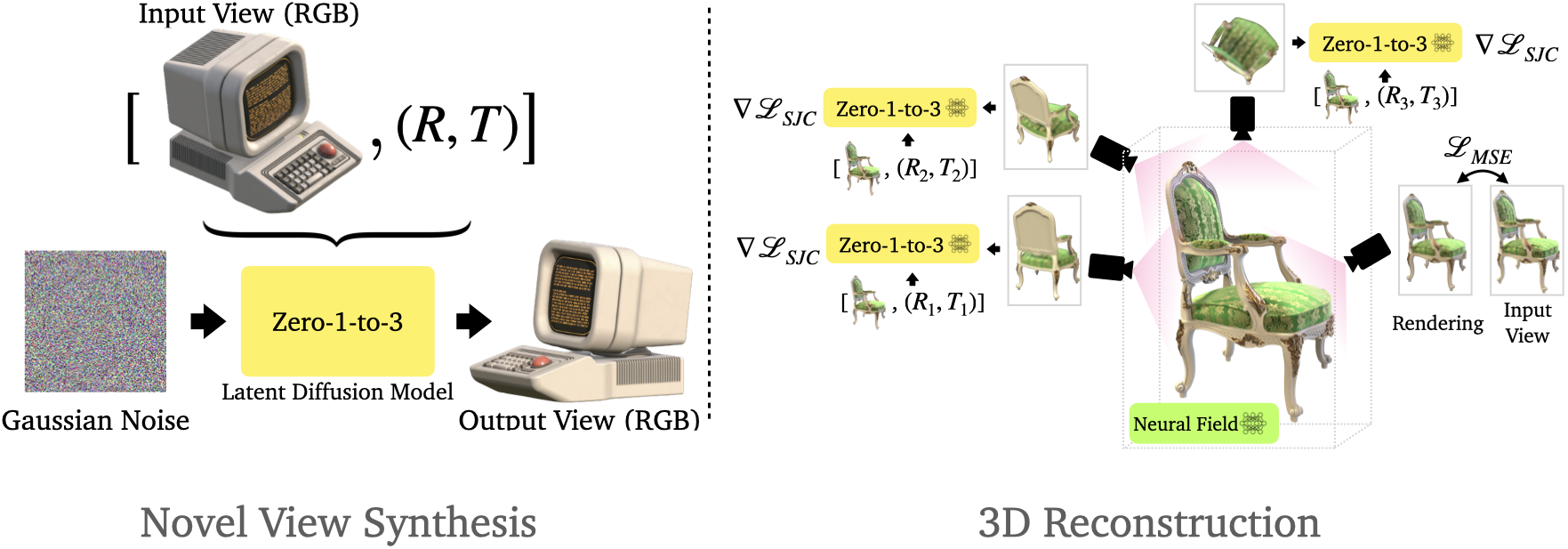}
	\caption{We show the pipeline of method Zero-1-to-3~\cite{liu2023zero}.}
	\label{fig:01}
\end{figure}

\begin{figure}[t]
	\centering

	\includegraphics[width=\linewidth]{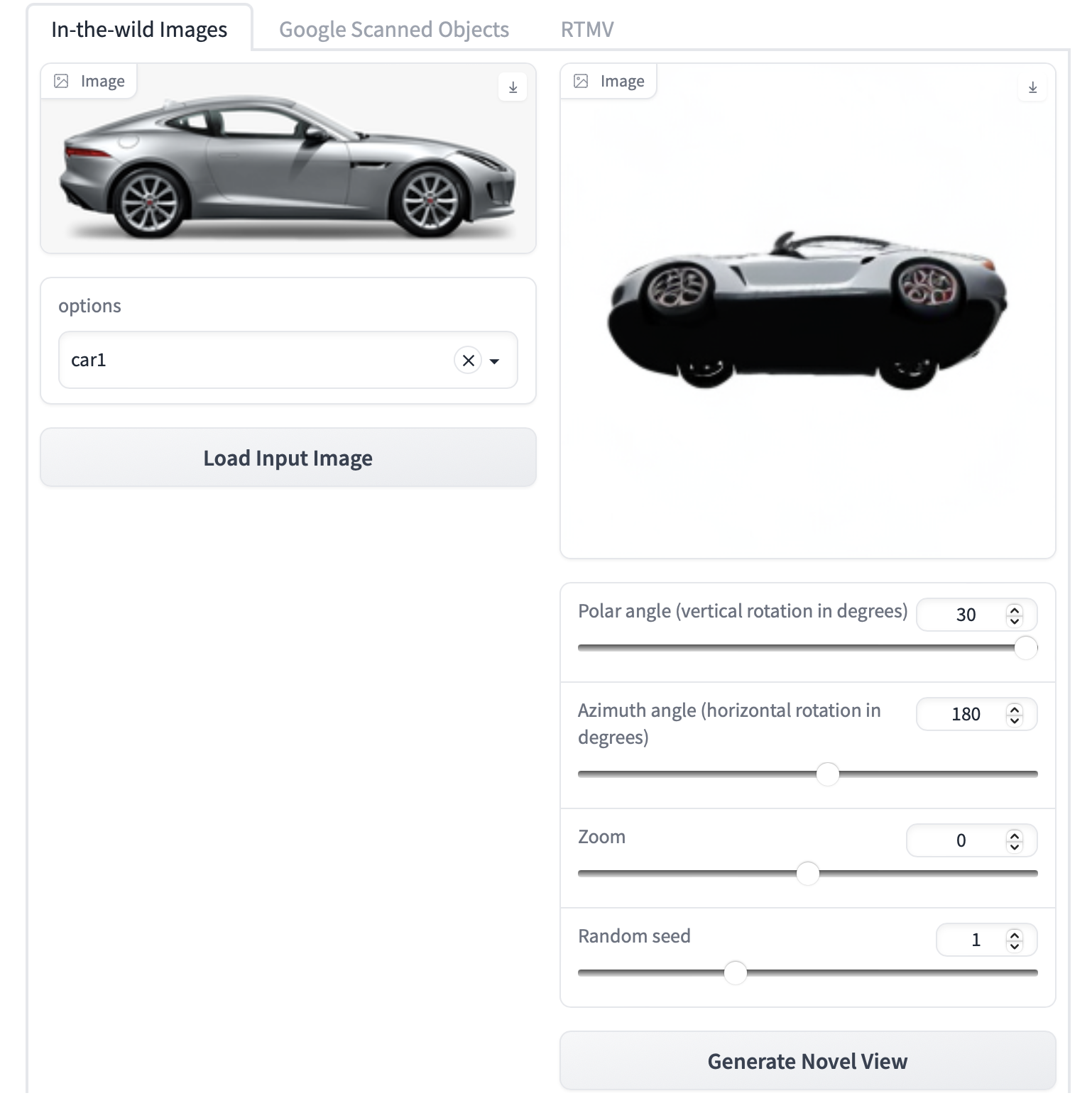}

	\caption{Zero-1-to-3 provides a visual template for users to easily create desired 3D assets according to their specific needs.}
	\label{fig:02}
\end{figure}

\section{Applications}

 This section explores the practical application of 3D generation technologies, highlighting their impact on different domains and their potential to revolutionize traditional workflows.

\subsubsection{Film and Entertainment Industry}

One of the most prominent applications of 3D generation is in the film and entertainment industry. Traditional methods of creating 3D assets for movies and visual effects often involve manual labor and extensive time investment. However, with advancements in AI-driven 3D generation, filmmakers can now streamline the production process by automating the creation of complex 3D models, characters, and environments. This not only accelerates the production timeline but also enables filmmakers to unleash their creativity without being constrained by technical limitations.

\subsubsection{Architectural Design and Visualization}
Architects and designers rely heavily on 3D visualization to convey their concepts to clients and stakeholders. AI-powered 3D generation tools offer architects the ability to quickly generate realistic architectural models from simple sketches or textual descriptions. By automating the tedious task of 3D modeling, architects can focus more on design iterations and exploring creative ideas. Additionally, these tools facilitate better communication between designers and clients, leading to more efficient decision-making processes.

\subsection{Virtual Reality (VR) and Augmented Reality (AR)}

VR and AR technologies have gained significant traction in recent years, offering immersive experiences in gaming, education, training, and simulation. AI-driven 3D generation plays a crucial role in enhancing the realism and interactivity of VR and AR environments. By dynamically generating 3D assets based on user interactions or environmental conditions, developers can create more dynamic and responsive virtual worlds. This not only enhances user engagement but also opens up new possibilities for interactive storytelling and experiential learning.

\subsection{Product Design and E-Commerce}

In the realm of product design and e-commerce, 3D generation technologies are revolutionizing the way products are conceptualized, prototyped, and marketed. By automatically generating 3D models of products from 2D images or textual descriptions, designers and retailers can showcase their products in a more immersive and interactive manner. This enables consumers to visualize products more accurately before making purchasing decisions, leading to higher customer satisfaction and reduced product return rates.

\subsection{Medical Imaging and Simulation}

In the field of medicine, AI-driven 3D generation has the potential to revolutionize medical imaging and simulation. By automatically converting medical scans such as MRI or CT images into detailed 3D models, healthcare professionals can better visualize and analyze complex anatomical structures. This facilitates more accurate diagnosis, treatment planning, and surgical simulations, ultimately leading to improved patient outcomes and reduced healthcare costs.

The applications of AI-driven 3D generation are vast and diverse, spanning across industries such as film and entertainment, architecture, virtual reality, product design, and healthcare. As technology continues to advance, we can expect these applications to become even more widespread, transforming the way we create, interact with, and experience 3D content in the digital age.

\section{Conclusion}
In conclusion, this paper has explored the emerging field of 3D content generation in the context of Artificial Intelligence Generated Content (AIGC) and large-scale models. While significant progress has been made, challenges such as data scarcity and computational limitations persist. However, leveraging pre-trained diffusion models shows promise in enhancing the efficiency of 3D content creation. By analyzing various methods and discussing challenges and potential solutions, this research provides insights into the current state and future directions of 3D content generation. Moving forward, further research is needed to overcome existing limitations and fully unleash the potential of AI in this domain.

\bibliographystyle{citation}
\bibliography{citation}

\end{document}